\documentclass[letterpaper,amsmath,amssymb]{revtex4}

\usepackage{graphicx}
\usepackage{dcolumn}
\usepackage{bm}
\usepackage{textcomp}
\usepackage{color}

\newcommand{\betsgacl}{$\lambda$-(BETS)$_2$GaCl$_4$}

\newcommand{\IV}{\hbox{I-V}}

\begin{document}


\title{Improvements to the Tunnel Diode Oscillator technique for high frequencies and pulsed magnetic fields with digital acquisition}

\author{William A.\ Coniglio}
\author{Laurel E.\ Winter}%
\author{Chris Rea}%
\author{Kyuil Cho}%
\author{C.\,C.\ Agosta}%
\affiliation{Clark University}%

\date{26 March 2010}

\begin{abstract}
We discuss improvements to the short-term performance of tunnel diode oscillator transducers with an emphasis on frequencies from 30\,MHz to 1.2\,GHz using LC type tank circuits. We specifically consider the TDO in pulsed high magnetic fields with fast digital acquisition. Since overdriven oscillators are necessary in pulsed fields to maintain oscillations, we examine the circuit using SPICE simulation during the design process and to gain insight into its behavior. We also discuss a numerical technique for demodulating the oscillations into frequency and amplitude.
\end{abstract}

\pacs{07.55.-w, 07.05.Kf, 84.37.+q, 84.30.Qi, 74.25.N-}
\maketitle

\section{Tunnel diodes and the TDO}
The tunnel diode is an excellent choice for an amplifier element in cryogenic circuits. \cite{clover_rsi_1970,vandegrift1975,brissontltdo,gevorgyan_rsi_2000,coffey2000} Because the operation of the diode is a consequence of quantum mechanics, it is resilient under a wide range of harsh conditions that do not interfere with the tunneling itself. Tunnel diodes can be used at frequencies well into the GHz, are stable at cryogenic temperatures, operate in magnetic fields of at least 45\,T, and can withstand significant amounts of ionizing radiation.\cite{dowdey_tdo_radiation} Disadvantages include the need to bias the 2-terminal device, sensitivity to heat (soldering) and static electricity, and the narrow range of different diodes presently available.

\begin{figure}\includegraphics[width=\columnwidth]{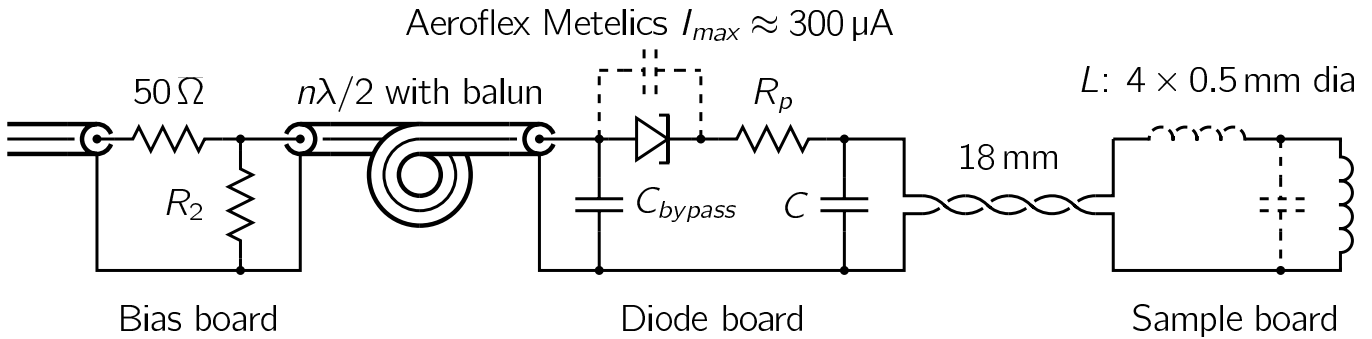}\caption{\label{fig:tdocircuit}Tunnel diode oscillator circuit used in 390\,MHz examination of the organic superconductor \betsgacl. $R_2=166\,\Omega$, $C_{bypass}=471$\,pF, $R_p=100\,\Omega$, $C=5$\,pF. The length of the tuned coax cable is 511\,mm.}\end{figure}

The basic Tunnel Diode Oscillator (TDO) consists of the diode as an amplifier element, a parallel LC tank circuit as the feedback element, and a bias network to provide the necessary DC bias as well as to couple small signals out of the circuit. In practical circuits, parasitic and distributed effects increase with frequency, and care must be taken to design the circuit properly. Our 390\,MHz oscillator from a superconductivity experiment\cite{coniglio_bets_2010} on the organic conductor \betsgacl\ appears in Fig.~\ref{fig:tdocircuit}.

Measurement circuits with a TDO may use either the capacitor or inductor as the transducer element. Either way, the oscillation frequency is perturbed by the variable component while the other remains constant. When a sample is placed inside the inductor, several parameters may be measured. With magnetic materials, the inductance will change in response to susceptibility. In conductors, the inductance responds to the RF skin depth, which is described by
\begin{equation}\label{eq:skindepth}\delta=\sqrt{\frac{\rho}{\pi f\mu}}\,,\end{equation}
where $\rho$ is resistivity, $f$ is frequency, and $\mu$ is the permeability ($\mu_0$ for most conductors). In a superconductor, magnetic field (DC and RF) is excluded from the center of the sample according to the London penetration depth ($\lambda$), although the presence of vortices complicates the measurement.\cite{coffey_clem_prl_1991} When the capacitor is the transducer and inductor is constant, measurable parameters include the dielectric constant between plates and distance or proximity if one plate is moveable. In cavity oscillators, the effective inductance and capacitance may both vary.

We have examined the TDO in relation to superconducting measurements in pulsed magnetic fields, which present unique challenges. The fields themselves are high (presently $\le50$\,T in our system), and magnetoresistance may play a role we still do not understand fully in the $Q$ of the resonant circuit. In addition to the size of the field, the rise rate, $dB/dt$ is also large ($\le7500$\,T/s for us). Eddy currents in any conductive materials within such fast fields heat significantly at cryogenic temperatures, and ground planes must be designed carefully and be as small as practical. Often, circuitry must be placed some distance from the sample to minimize heating. On the positive side, temperature stability is not a major concern, because the time scale of the measurement is so short. Most of the improvements discussed below are specific to fast time scales and pulsed fields, although some still apply over longer time scales.

\section{Harmonic behavior}
In the harsh environment of pulsed fields, marginal (small-signal) oscillators are unreliable. Magnetoresistance and electromagnetically induced changes in the diode bias will upset a carefully tuned circuit. When oscillations are larger than marginal, nonlinear effects create harmonic currents through the diode.

Often, the sample must be mounted on a rotating platform or otherwise placed some distance from the diode. Wires connecting portions of the circuit should be treated as transmission lines. A tank circuit formed with lengths of transmission line and an LC circuit will have a spectrum of resonant frequencies, rather than a single one. The spectrum of a resonant cavity with an inductor at one end is treated in \cite{brissontltdo,catalinmanual}. It should be noted that due to the tremendous nonlinearity of the diode, simultaneous excitation of non-harmonically related modes is not possible. Due to parasitic capacitances, the oscillator will generally collapse into the lowest available frequency, although a sufficiently high impedance mode with $2\pi n$ phase shift can be made to oscillate. Any harmonically unrelated frequencies that are excited will be incoherent over the period of the main oscillator frequency and their power will become buried in amplitude-modulated harmonics of the fundamental.

Harmonics of the fundamental frequency are also generated by nonlinearities in the \IV\ curve of the diode and appear as current through the diode. Since we measure the voltage across the bypass capacitor $C_b$, the measured voltage at a particular frequency is proportional to the diode current and contains these harmonics. A parallel tuned LC circuit nominally resonates at only one frequency, and harmonic currents pass through the tank without dissipating power, but those currents do produce a voltage across $C_b$, so the output signal from the TDO contains harmonics not radiated by the coil. In addition, since the fundamental signal measured across $C_b$ is proportional to oscillating current and not voltage, its amplitude is a complicated function of the circuit, diode characteristics, and the sample, which makes it unreliable as a measurement parameter.

\section{Spice simulation}
\begin{figure}\includegraphics[width=\columnwidth]{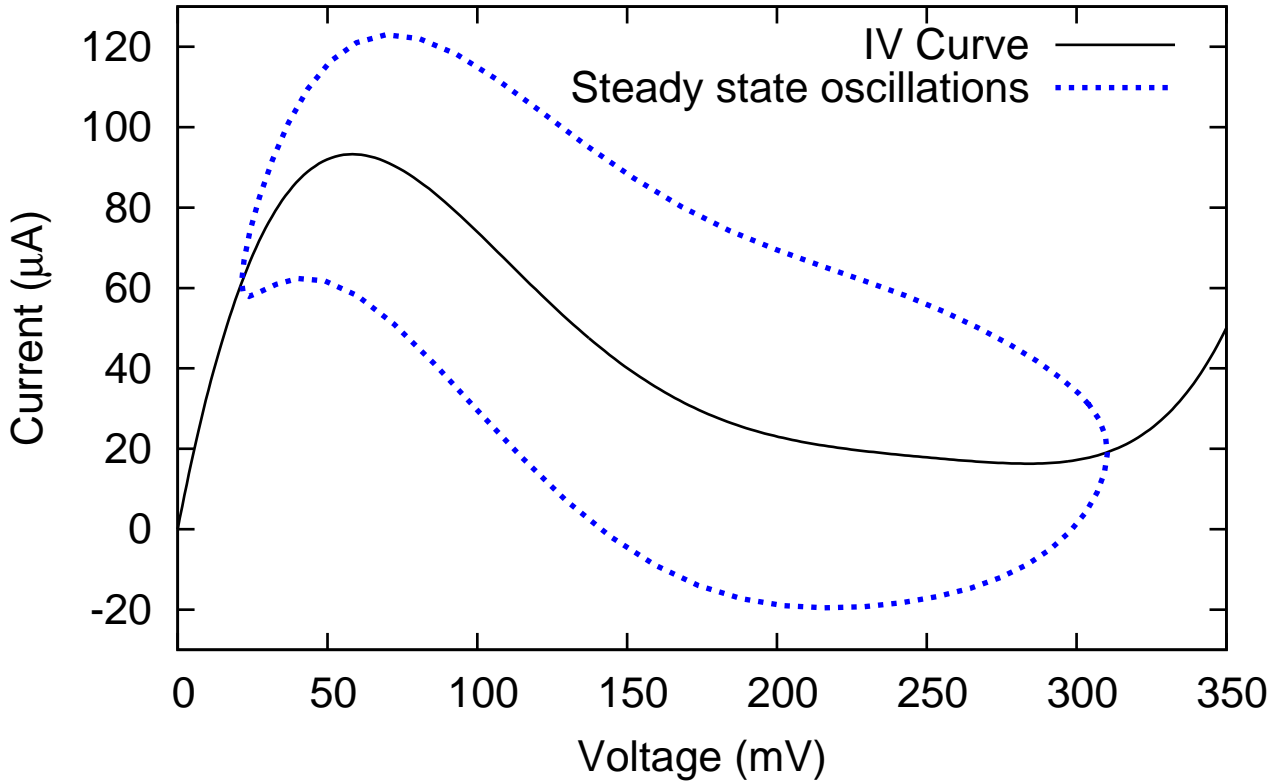}
\caption{\label{fig:ivcurve}Measured current vs.\ voltage curve of a tunnel diode (solid) and simulated oscillations. Hysteresis between current and voltage arises due to the parasitic and junction capacitances in the diode. As the frequency rises, capacitive effects become more pronounced, capping the maximum useable frequency of the device.}
\end{figure}

Calculation techniques\cite{vandegrift1981,gevorgian_rsi_1998} must make assumptions that simplify the circuit and lend themselves to exact solutions or approximation techniques. Traditional assumptions about the TDO are that circuit components are very close together, components behave ideally with respect to frequency, and that the oscillations are linear or only slightly overdriven. High frequency oscillations with elements connected by lengths of transmission lines and overdriven for use in pulsed fields violate all of these conditions. Following good RF design practice\cite{horowitzhill,diefenderfer,ludwigbretchko}, we model our circuits in SPICE\cite{ngspice,reathesis2006} and make adjustments before building cryogenic versions. Fig.~\ref{fig:ivcurve} shows an example \IV\ curve and the simulated oscillations of the current and voltage across the pins of the diode. The small ($\approx$0.3\,pF, dependent on voltage) capacitance within the diode rotates the current and voltage away from perfect in-phase behavior, and figures significantly in the output amplitude, which is proportional to diode current.

We modeled the diode as a voltage controlled current source in parallel with a small capacitor. We have attempted to account for parasitic capacitances and inductances throughout the circuit individually.

\section{Experiments to optimize short term noise performance}
To date, guidelines for minimizing noise in the TDO have focused on the long-term performance of the oscillator. Since our experiments are performed in pulsed fields that rise in tens of milliseconds and fall in a few hundred milliseconds, we are primarily concerned with phase noise (sub-microsecond jitter in phase or frequency), which will average out over longer time scales. To study the circuit under simple conditions, we constructed a series of 12\,MHz oscillators with a number of different component choices, all operating at room temperature. We also tested our circuits at a range of bias voltages. To measure phase noise, we heterodyned our signals down to 500\,kHz, then digitized them. Our Fourier demodulation method discussed later in this article extracted the frequency as a function of time with an averaging time of 18\,\textmu s. We computed the average frequency and standard deviation as our measure of noise.

We found that short term (phase) noise was lowest under the following circumstances. For each observation, I will suggest a mechanism:

\begin{figure}\includegraphics[width=\columnwidth]{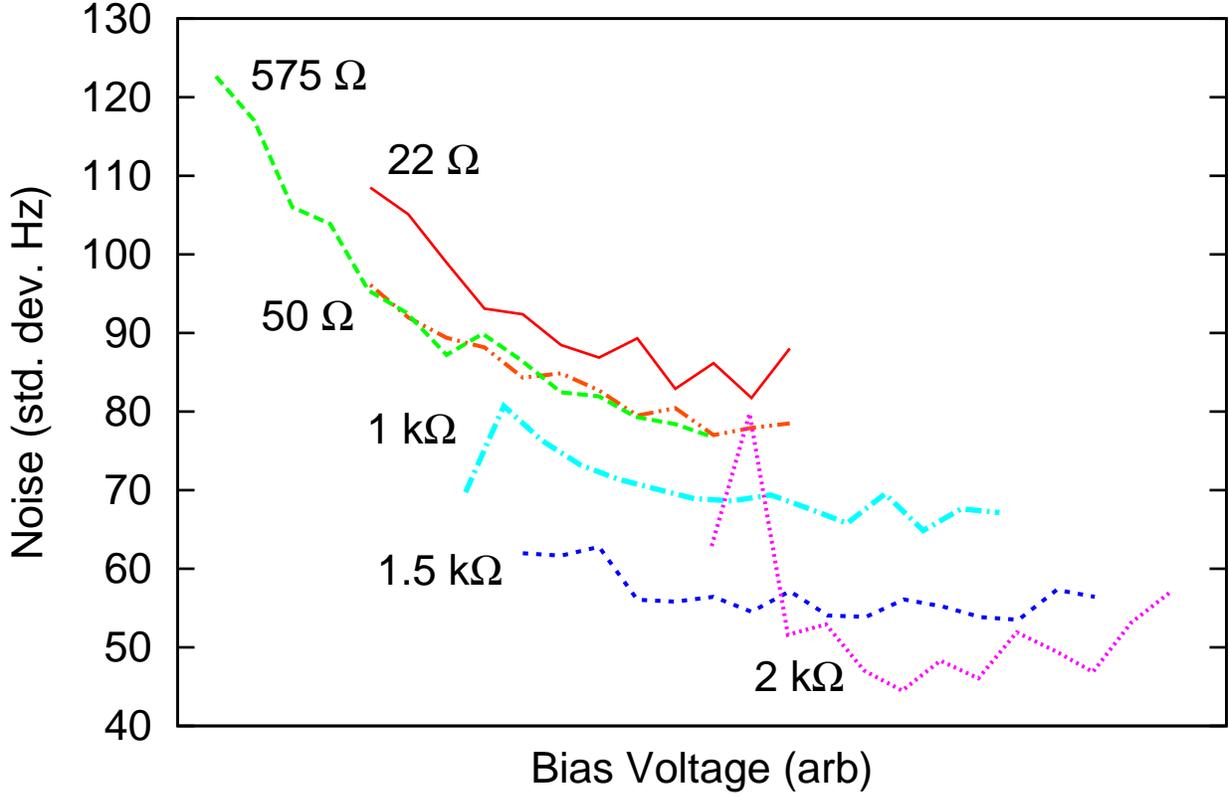}
\caption{\label{fig:rpnoise}Frequency noise vs.\ bias voltage using different values for the parasitic resistor. Tunnel diode $\vert R_n \vert\approx300\,\Omega$, and oscillator frequency $f_0\approx 12\,\textnormal{MHz}$. Noise was determined using the digital demodulation technique discussed below with an averaging time of 18\,\textmu s. We took several milliseconds of data, then computed the standard deviation of all points to quantify short-term noise.}
\end{figure}
\subsection{Higher diode bias voltages}
We found that phase noise decreases with higher bias voltage until the circuit stops oscillating. (See Fig. \ref{fig:rpnoise}) Since the sharpest nonlinearity in the \IV\ curve is the peak current region of the diode bias, we suspect that higher voltages keep the oscillation center further from the \IV\ peak and result in fewer unnecessary harmonics. The effective negative resistance of the diode also decreases (becomes more negative) with greater bias voltage, so the circuit may be bias-tuned for more marginal oscillation. In a sense, this tuning is a form of matching the diode to the tank impedance, rather than the other way around, as recommended by \cite{vandegrift1975,brissontltdo,gevorgyan_rsi_2000}. Bias-tuning is only a fine technique, however. Should experimental conditions allow, it may be beneficial to use the combination of an approximately tapped coil or transmission line and diode bias fine-tuning. In pulsed fields, where the coil is some distance from the oscillator and mounted on a rotating platform, we have not developed a technique for tapping the coil. For reliability in pulsed fields, we find it best to stay well within the oscillating range of bias.
\subsection{Increase $R_p$}
As also depicted in Fig.~\ref{fig:rpnoise}, the larger parasitic resistors resulted in less noise. While the fundamental oscillations are dissipated both in the tank circuit and parasitic resistor $R_p$, harmonics caused by the diode are dissipated only in $R_p$\,. Since power dissipation for harmonics is $I^2R_p$, the larger resistors dissipate more of the energy created by overdriven harmonics in the diode.
\subsection{Adjust $L/C$ to maximize $Q$}
\begin{figure}\includegraphics{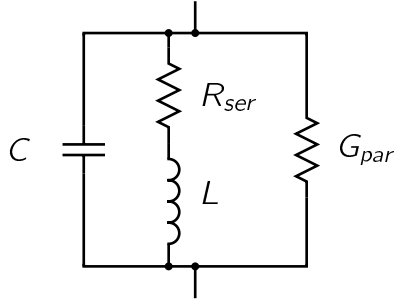}\caption{\label{fig:tankcircuit}Tank circuit}\end{figure}
We observed that the fractional noise decreased as we increased the capacitance in the tank circuit. A possible explanation is that we were simply increasing the quality factor ($Q$) of the circuit. The overall effect of $L/C$ on the resonant circuit quality factor deserves attention.

We model our tank circuit as shown in Fig.~\ref{fig:tankcircuit}. $R_{ser}$ is the series resistance of the wires, and $G_{par}$ is a fictitious conductance used to represent RF radiation and dissipation. It is assumed that $R_{ser}$ and $G_{par}$ are small. Note that conductance is the inverse of resistance, so $G = 1/R$. The quality factor of the circuit in Fig~\ref{fig:tankcircuit} is
\begin{equation}\label{eq:qtank}Q=\sqrt{\frac{L}{C}}\left(\frac{1}{R_{ser}+G_{par}L/C}\right)\,.\end{equation}
Assuming the design frequency is fixed, we must optimize $L/C$ for maximum $Q$. If on one hand, the wire resistance term $R_{ser}$ dominates, we will do best to increase $L$. On the other hand, if the radiation/dissipation term involving $G_{par}$ is larger, we should increase the capacitance. To show that this is the case, it is easy to take the differential of $1/Q$ with respect to $L$ and $C$. (The differential of $Q$ itself is harder.) We find that $1/Q$ is a minimum (and therefore $Q$ is maximum) when $R_{ser} = G_{par}L/C$.

Since the relevant parameters are not easy to measure without a network analyzer, and may even change depending on experimental conditions, we must rely on a observation to determine which limit applies at any given time. In the \betsgacl\ experiment, we noticed that we could receive the oscillator signal by tuning a narrow-band FM radio to the appropriate frequency. With a sample in the coil, the range was about 8 feet through a wall. Without the sample, we could still detect the signal after it passed through another wall. In the case of the sample and coil, dissipation clearly plays a role. On the other hand, the empty coil emitted detectable RF, so we expect that even without a sample, $G_{par}$ cannot be ignored.

In order to better understand the dependence of the quality factor of our circuit on $R_{ser}$ and $G_{par}$, we simulated the circuit in SPICE. We modeled the diode as a polynomial voltage controlled current source with appropriate parasitic components in series and parallel. We compared our simulation with the actual circuit by sweeping the bias voltage and matching the simulated oscillation range with the actual measurement for different values of $G_{par}$ (while holding $R_{ser}=0$). From this exercise, ($L=24\,$nH, $C=7\,$pF) we estimate $Q\approx10$, which suggests that $R_{ser}+(3400\,\textnormal{H}/\textnormal{F})G_{par} \approx 5.8$. Since the tank circuit operated at cryogenic temperatures, and all wire connections were made of copper, it is doubtful that even an ohm could be attributed to series resistance in the absence of magnetoresistance. We conclude that for low field work, the quality factor is dominated by the $G_{par}$ radiation and dissipation term, and $Q$ may usually be increased using larger capacitors and smaller inductors. As frequencies (and skin depths) increase and magnetoresistance appears at high fields, we become less confident with this analysis. Further study of tiny wirewound inductors in high magnetic fields is necessary to fully understand the $Q$ dependence.

\section{Distributed component considerations}
We designed the circuit in Fig.~\ref{fig:tdocircuit} for a penetration depth study of \betsgacl. It consists of three lumped circuits with transmission lines between them. This so-called split-circuit design has a number of advantages. It allows us to choose the (hopefully small) distance from the sample coil to the oscillator circuit. Any distance between the sample and diode reduces the oscillation frequency and provides opportunities for sources of interference or dissipation. Mounting the bias network halfway up the probe allows $R_2$ to be a stiff voltage divider without adding extra heat near the sample. The distributed design also allows for flexibility and modularity in design, construction, and troubleshooting. We found time domain reflectometry to be a useful technique for troubleshooting the probe \textit{in situ}, as we could easily pinpoint the location of electrical irregularities.

After testing the basic design in liquid nitrogen to tune the operating frequency, we constructed the three boards. The sample and coil were mounted to a one-axis rotating platform at the bottom of the probe. The diode board sat nearby with its ground plane perpendicular to the applied field to minimize eddy currents. The bias network was attached about halfway up the probe, and was located above the liquid $^3$He level. The lower section of the probe was made of G-10 plastic, and the upper portion was stainless steel. We secured the coax and ground plane of the upper board to the stainless steel portion of the probe. All metal parts of the probe were solidly grounded to each other and to the RF coax, but ground shields for all other signals elsewhere in the room were isolated from the probe and grounded by a different path. The probe top was grounded directly to the central grounding point of the lab with copper braid. Ferrite beads around each cable end and intermittently on long runs protected the grounds from becoming antennas.

We made substantial use of the complex reflection coefficient $\rho=(Z/Z_0+1)(Z/Z_0-1)$ and Smith Charts to optimize our designs. The transmission line from the oscillator board to the coil affects the oscillation frequency and has been treated carefully by \cite{brissontltdo,catalinmanual}. The transmission line between the oscillator and matching network requires attention as well. The bypass capacitor $C_b$ is made to be large, so it acts as a nearly short circuit and does not perturb the oscillator. Because of the short circuit condition, the wave entering the coax is almost entirely made of current, or $\rho = -1$. Rotating the wave through a cable length that is any integer multiple of a half wavelength ($\ell = n\lambda/2$), the wave will again be carried by the current. For long wavelengths and short cables ($\ell<\lambda/10$), the transmission line is unimportant, and $n=0$. Since the voltage across $R_2$ is nearly zero, it does not figure into the AC analysis. The 50\,$\Omega$ resistor matches the signal to the coaxial cable leading out of the probe. In addition, the matching network ensures that all signals traveling down the probe will be conditioned to reflect off $C_b$ at high-SWR.

We suspect that the ground plane of the diode board is a significant source of heat at low temperatures, and we made it as small as we thought practical. We found that in some circuits, high SWR and unbalanced conditions caused unwanted radiation from the ground. To guard against common-mode currents, we coiled a five-turn current balun\cite{arrl1989} (depicted schematically by a coax loop in Fig.~\ref{fig:tdocircuit}). Since we had taken steps already to eliminate high SWR, the balun was a precaution, and we did not evaluate its performance. We have since conducted an experiment using a similar circuit with no current balun and see no ill effects in its absence.

It appears that we have succeeded in stabilizing the circuit against standing waves and poor grounding, as we have been able to take measurements simultaneously with two TDOs (one from Fig.~\ref{fig:tdocircuit} and one from Fig.~\ref{fig:tdoampcorrect}) on the same probe with no interference between them.

\section{Amplitude-sensitive variations}
\begin{figure}\includegraphics[width=\columnwidth]{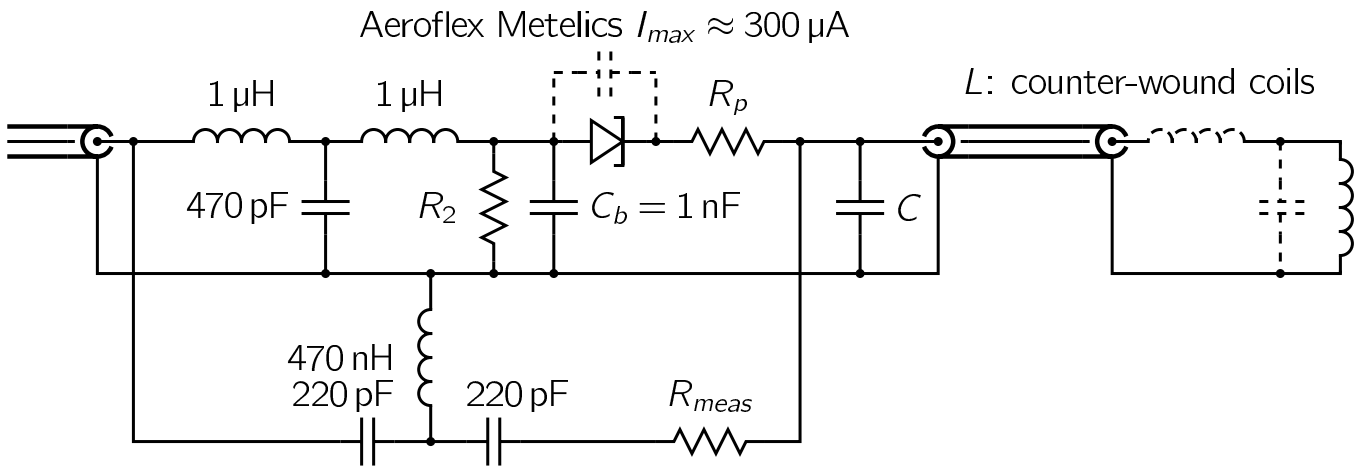}\caption{\label{fig:tdoampcorrect}TDO design with an output proportional to the voltage across the tank circuit rather than the current. $R_{meas}$ (2--10\,k$\Omega$) forms a passive probe and voltage divider with the 50\,$\Omega$ termination of the coax cable. High-pass (15\,MHz) and low-pass (4\,MHz) filters diplex the DC input with the RF output for use at the end of a single coax on a probe.}\end{figure}
The split-circuit TDO described above is not without its drawbacks. While it accommodates UHF wavelengths, a small form-factor, and the capability to run in environments that require very low heat dissipation at the diode board, it does not deliver a useful amplitude measurement. For applications where it is necessary or desirable to measure the voltage across the resonant tank, the TDO must be modified to accommodate the passive measurement while maintaining the DC characteristic to bias the diode.

A diagram of the circuit is shown in Fig.~\ref{fig:tdoampcorrect}. The diode, tank circuit, $R_p$, and $C_b$ serve the same functions as before, but instead of coupling the signal out through capacitor $C_b$, a new resistor, $R_{meas}$, plays the role of a passive (transmission-line) probe and acts as the input arm of a voltage divider. The $50\,\Omega$ transmission line leading up the probe and the $50\,\Omega$ input impedance of the receiver circuitry form the other arm of the voltage divider. The input impedance of the first amplifier (or preselector, if used) must match the characteristic impedance of the transmission line, or standing waves will develop. The inductors and capacitors at the left of the diagram form a diplexer to allow the DC bias voltage to reach the diode while coupling the RF signal out of the circuit. Inductors should be chosen with self-resonant frequencies as high as possible, but due to the passive impedance matching, they may operate above their self-resonant point with no ill effects.

\section{High bandwidth digital demodulation}
Using analog techniques, we first heterodyne our signals down to $5\pm2$\,MHz for digital acquisition at 10\,MS/sec. In software, we use a technique we have developed for demodulating the digitized signal into its frequency and amplitude. The frequency and amplitude are related to the London penetration depth and dissipation in superconducting samples and the skin depth in metals. The algorithm employs a spectral approach locally and is repeated at intervals over the length of the data, which allows for a computationally parallel approach.
\subsection{Other approaches}
In the past, we have tried various analog and digital approaches to demodulating the TDO signal. Specifically, we have considered the analog discriminator described in \cite{coffey2000} and a digital technique that locates each peak of the wave by fitting a function. Both of these techniques work well at frequencies below 1\,MHz, but are insufficient for recent wide band work. We instead turn to digital realizations of spectral techniques.

The Fast Fourier Transform (FFT) is a popular choice for spectral analysis because of its speed. The decimation technique that speeds up the FFT also makes it useless if the exact frequency basis is important but unknown. We require frequency precision only attainable from an FFT by zero-padding the end of the data up to some large size. In any case, the off-basis frequencies are aliased onto nearby frequencies, broadening the spectrum. We are also interested in only a small portion of the spectrum; the FFT does far more computations than are necessary.

We have developed a technique to demodulate the frequency by taking individual Fourier integrals (not FFT) at chosen frequencies iteratively until we locate the frequency of the maximum.
\subsection{Choice of window}
We look at only a small portion of our data at a time through use of a window function. The choice of window determines the bandwidth and sidebands to which we are sensitive. We chose a raised cosine window because it has finite impulse width, small side-lobes, and weights each ADC point equally. If side-lobes were of paramount concern, a Gaussian could be approximated to some larger impulse, but the sum over the set of all windows would favor some data over others. For a raised cosine window of length $T$, demodulated points should be computed at intervals of $T/2$.
\subsection{Maximizing Fourier integral coefficient}
After windowing the input to create a wave packet $\Phi(t)$, we numerically maximize the Fourier integral with respect to frequency to get the peak amplitude,
\begin{equation}A[\Phi(t)] = \textnormal{max}_f\left(\left\|2\pi\kern -4pt\int\kern -4pt\Phi(t)e^{2\pi ift}\,dt\right\|^2\right)\,.\end{equation}
\subsection{Conditioning and computational optimizations}
We wrote the algorithm in portable C++ with a command line interface. We used OpenMP for multithreading, and modified the Template Numerical Toolkit (TNT)\cite{nist_tnt} to make it thread-safe. We also wrote and utilized a C++/TNT library for data analysis that we based on some of the excellent ideas in \cite{numericalrecipes}.

The algorithm consists of two stages. Stage~1 demodulates the points independently, with one output point corresponding to each placement of a window on the original data. Stage~2 ensures the output is phase consistent with the original oscillations.

There are a number of tricks to speed up and stabilize the Stage~1 searching algorithm. Since the computations are independent, we parallelize this portion of our algorithm. As much as possible, we perform arithmetic with 64- or 128-bit fixed point (integer) techniques rather than floating point. Computation of each output point begins with a FFT to locate the vicinity of the peak and form a frequency bracket, within which we perform our search. We use Brent's Method\cite{numericalrecipes} on the bracket to search for the peak and return the frequency, amplitude, and phase information.

Once the entire data set is demodulated, Stage~2 begins by comparing the phase information with the expected phase based on frequency and point spacing to produce a frequency error data set. (Error magnitude for our data was less than 0.2\% of the total frequency and occurs when the frequency is changing rapidly.) We also verify that the signal is approximately phase continuous. Since the error data includes significantly more noise than the uncorrected frequency, we smooth our error data (analogously to a PLL low-pass filtering the phase error signal) then apply it back to the original frequency. The final frequency set is phase continuous, meaning an integral of the frequency over an interval will equal the number of periods (including fractions of a period) in that interval. Phase discontinuous data, such as the beginning of a magnetic field pulse, are thrown out, leaving downstream analysis techniques to decide what to do with the gaps.

\subsection{Performance}
After careful optimization, we have combined numeric precision with computational speed that is acceptable for use in pulsed field experiments. Precision is limited by the quality of input data. A computer-generated sine wave in double-precision demodulates (with a window size of 512 points and effective averaging width of 256 points) into a frequency and amplitude with a standard deviation error of $7.5\times10^{-11}F_N$ where $F_N$ is the Nyquist frequency (half the sampling rate). The same sine wave at 16-bit integer precision gives $1.7\times10^{-9}F_N$, and with 8 bits of data, $5\times10^{-7}F_N$. A sampling of pre-shot TDO data from the \betsgacl\ experiment suggests our experimental noise standard deviations were about $2.7\times10^{-5}F_N$, (270\,Hz) or about 17\,dB above the demodulation noise. (Note: compared to the free-running oscillator frequency of 390\,MHz, our 270\,Hz noise represents a $\pm0.7$\,ppm measurement every 12.8\,\textmu s!)

Running two processor cores, we can demodulate approximately one million samples per second. Because the algorithm makes heavy use of wide integer math, we noticed a significant benefit when compiling the code in 64-bit for capable architectures.

Interestingly, this technique simultaneously demodulates the amplitude and frequency from the data. With a generalization of the windowing function to an appropriate operator, other demodulations may be performed with a similar mechanism.

\section{Acknowledgements}
We acknowledge support from the U.\,S. Department of Energy ER46214.

\nocite{srikanth_rsi_1999}
\nocite{ohmichi_komatsu_rsi_2004}
\nocite{altarawneh_mielke_brooks_rsi_2009}
\bibliography{bibs/tdo,bibs/computation,bibs/bets}{}

\end{document}